# Conditional quantum distinguishability and pure quantum communication

Tian-Hai Zeng

*Department of Physics, Beijing Institute of Technology, Beijing 100081, China*

I design a simple way of distinguishing non-orthogonal quantum states with perfect reliability using only quantum CNOT gates in one condition. In this way, we can implement pure quantum communication in directly sending classical information, Ekert's quantum cryptography and quantum teleportation without the help of classical communications channel.

Indistinguishability of non-orthogonal quantum states is at the heart of quantum computation and quantum information[1]. Although it is possible for distinguishing non-orthogonal quantum states some of the time[1-3], the proof[1] and no-cloning theorem[4] that we cannot exactly clone an unknown quantum state prohibit distinguishing non-orthogonal quantum states with perfect reliability.

So far, all quantum communications that can be implemented have to be helped with classical communications channel[1] (CCC) in which information can be sent by electromagnetic wave or broadcast, or sending quantum system (or qubit). For examples, suppose Alice communicates with Bob, quantum cryptography or quantum key distribution[5-9] includes CCC used in communicating the information of choice of measurement bases between Alice and Bob, and used in sending qubit. Quantum teleportation[10-13] includes CCC in which Alice must transmit her measurement result to Bob. Dense coding[14-16] includes CCC in which Alice sends qubit to Bob.

If a quantum communication can be implemented without the help of CCC, we call it *pure quantum communication* (PQC). It is indistinguishability of non-orthogonal quantum states that prevents PQC in Ekert's quantum cryptography[9] and the quantum teleportation[10-13] that need sending information but need no sending qubit. Sending qubit must be in CCC, therefore it prevents PQC in the quantum cryptography[5-8] and the Dense coding[14-16].

In this letter, I design a simple way of distinguishing non-orthogonal quantum states with perfect reliability using only quantum CNOT gates in one condition. In ref. 17, I have designed a way of doing same things using single qubit gates and quantum CNOT gates in the condition. So we can surpass the confine of the proof and no-cloning theorem and make use of the ways in PQC in directly sending classical information, Ekert's quantum cryptography[9] and the quantum teleportation[10-13] without the help of CCC.

Suppose Alice and Bob are separated in space. They share an EPR[18] pair $|\psi\rangle = (|00\rangle+|11\rangle)/\sqrt{2}$. Alice wants to send a bit classical information to Bob, she can measure her qubit by selecting one of the two bases: $\{|0\rangle, |1\rangle\}$, or $\{(|0\rangle\pm|1\rangle)/\sqrt{2}\}$. If Alice measures her qubit by selecting the first base, then Bob's qubit will be in state $|0\rangle$ or $|1\rangle$ instantaneously [1]. If Alice selects the other base, then Bob's qubit will be in state $(|0\rangle+|1\rangle)/\sqrt{2}$ or $(|0\rangle-|1\rangle)/\sqrt{2}$ instantaneously. For getting a bit classical information from Alice, Bob must distinguish the two set states $\{|0\rangle, |1\rangle\}$ and $\{(|0\rangle\pm|1\rangle)/\sqrt{2}\}$.

In the case of distinguishing two arbitrary non-orthogonal states, we should be given one condition that we know one of the two arbitrary non-orthogonal states being in eigenstate and the *direction* of the eigenvector. In communications, Bob knows the *directions* of two eigenvectors, and he can do projective measurements on his qubit on one *direction* of an eigenvector.

Here is the step. After Alice measures her qubit, Bob's qubit is in one of the four states instantaneously, Bob's task is to distinguish the two set states. Bob lets his qubit to control qubit line and an ancilla qubit to target qubit line of a quantum CONT gate (Fig. 1a). After operation of a gate, he does projective measurement on the output of target qubit on one *direction* of an eigenvector, then Bob lets the output of control qubit to control qubit line of second one, then does same thing again.

Next is the detail of the step. Fig.1b shows the quantum circuit. Bob selects each ancilla qubit (target qubit) being in the state $(|0\rangle+2|1\rangle)/\sqrt{5}$ (or $a|0\rangle+b|1\rangle$). If Bob's qubit is initially in state $|0\rangle$, he measures each target qubit passed through each gate in state $|0\rangle$ with probability 1/5. After 5 measurements, Bob gets one target qubit in state $|0\rangle$. If Bob's qubit is initially in state $|1\rangle$, then each target qubit passed through each gate will be in state $(2|0\rangle+|1\rangle)/\sqrt{5}$ or in state $|0\rangle$ with probability 4/5. After 5 measurements, Bob gets 4 target qubits in state $|0>$. So if Alice selects the base $\{|0>, |1>\}$, then Bob will get the number of target qubits in state $|0>$ tending to 1 or 4.

If Bob's qubit is initially in one of the two states $(|0> \pm |1>)/\sqrt{2}$, then the two output qubits will be in one of the two entangled states:

$$\{|0>(|0>+2|1>)/\sqrt{5} \pm |1>(|1>+2|0>)/\sqrt{5}\}/\sqrt{2} =$$
$$\{[(|0> \pm 2|1>)/\sqrt{5}]|0> + [(2|0> \pm |1>)/\sqrt{5}]|1>\}/\sqrt{2}. \qquad (1)$$

Bob measures the target qubit passed through the first gate in state $|0>$ with probability 1/2.

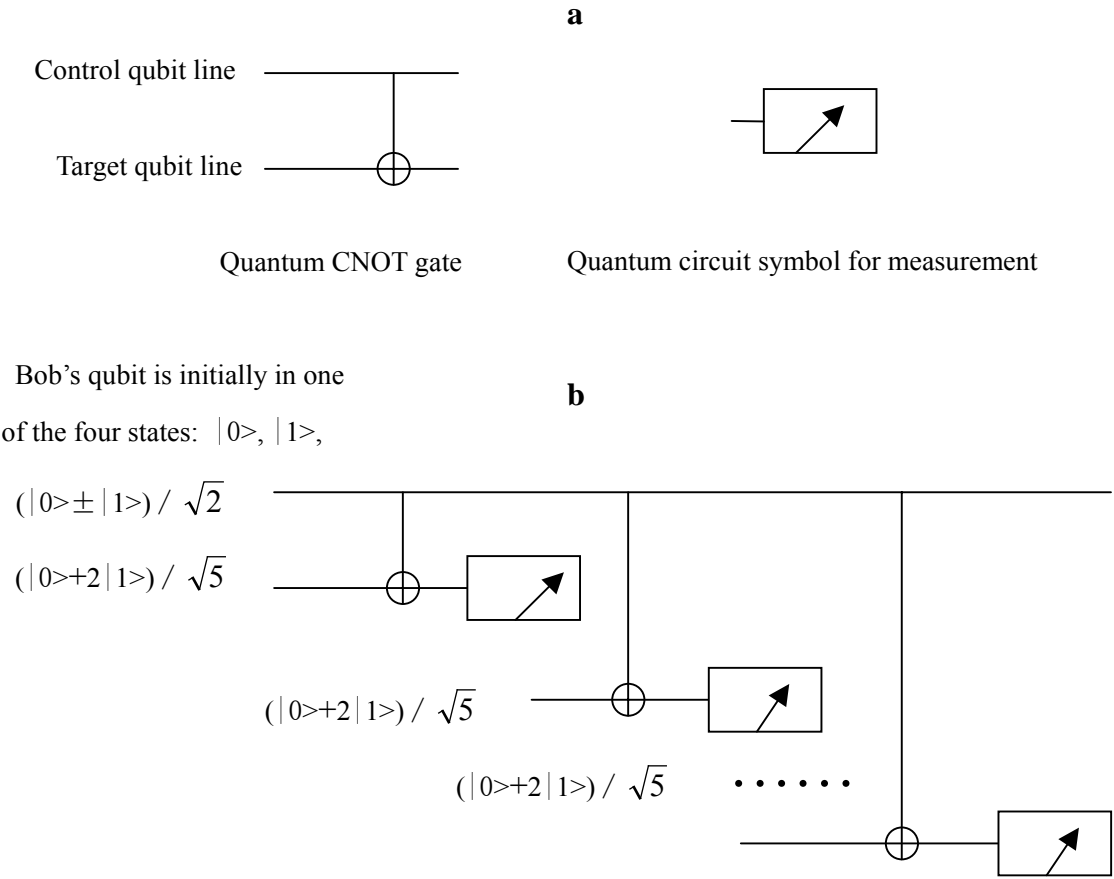

FIG. 1. Quantum CNOT gate and quantum circuit symbol for measurement (a) and quantum measurement circuit for Bob's qubit (b).

If Bob gets the first target qubit in state $|0>$, then the state of control qubit becomes one of the two states $(|0> \pm 2|1>)/\sqrt{5}$ from Eq.(1); otherwise becomes one of the two states

$(2|0>\pm|1>)/\sqrt{5}$. Bob lets the qubit in one of the four states $(|0>\pm2|1>)/\sqrt{5}$ and $(2|0>\pm|1>)/\sqrt{5}$ and the second ancilla qubit in state $(|0>+2|1>)/\sqrt{5}$ to the control qubit line and target qubit line of the second gate respectively. In the case of $(|0>\pm2|1>)/\sqrt{5}$, the two output qubits will be in one of the two entangled states:

$$\{|0>(|0>+2|1>)/\sqrt{5}\pm2|1>(|1>+2|0>)/\sqrt{5}\}/\sqrt{5}=$$
$$\{[(|0>\pm4|1>)/\sqrt{17}]\sqrt{17}|0>+[(|0>\pm|1>)/\sqrt{2}]2\sqrt{2}|1>\}/5. \qquad (2)$$

Bob measures the target qubit passed through the second gate in state $|0>$ with probability 17/25. In the case of $(2|0>\pm|1>)/\sqrt{5}$, the two output qubits will be in one of the two entangled states:

$$\{2|0>(|0>+2|1>)/\sqrt{5}\pm|1>(|1>+2|0>)/\sqrt{5}\}/\sqrt{5}=$$
$$\{[(|0>\pm|1>)/\sqrt{2}]2\sqrt{2}|0>+[(4|0>\pm|1>)/\sqrt{17}]\sqrt{17}|1>\}/5. \qquad (3)$$

Bob measures the target qubit passed through the second gate in state $|0>$ with probability 8/25. This method that only tests ancilla qubit has been used in quantum error-correction[1]. Intuition and calculations tell us that after 5 measurements, Bob gets that the number of target qubits in state $|0>$ *will not tend to* 1 or 4, but in the case that Bob's qubit is initially in one of the two states $|0>$ and $|1>$, he will get the number of target qubits in state $|0>$ tending to 1 or 4.

Although in this way, Bob still cannot distinguish the two set states with perfect reliability, Alice and Bob can use a group of EPR pairs to send a bit classical information, that is, Alice selects same base on the group of EPR pairs. So, Bob can get a group of numbers corresponding to the base Alice selected or one set states. Bob can do statistics on these numbers (or *second level statistics*) and get result that the statistical distribution of these numbers is *discrete* corresponding to the set states $\{(|0>\pm|1>)/\sqrt{2}\}$, or has *two convergent values* 1 and 4 corresponding to the set states $\{|0>,|1>\}$. Finally, he can make use of the difference to distinguish the two set states with perfect reliability and get a bit classical information.

Quantum communication can send both classical and quantum information. By use of this way, we can implement pure quantum communication in directly sending classical information, Ekert's quantum cryptography[9] and the quantum teleportation[10-13] without the help of classical communications channel.

Bell's inequality[19] continues to be examined[20]. If Bell's inequality can be proved and we can get long time EPR pairs against decoherence in future, pure quantum communication implies that sending information can be fast than light! We believe that this will excite much studies on many relative issues.

In summary, I design a simple way of distinguishing non-orthogonal quantum states with perfect reliability using only quantum CNOT gates in the condition. We emphasize that the key of distinguishing the two set states is the difference of discrete and two convergent values of the statistical distribution. We expect that the conditional quantum distinguishability will be proved in experiments and those pure quantum communications can be implemented.

I thank B. Shao, J. Zou and J. F. Cai for useful comments.